\documentclass[twoside]{article}
\usepackage{amsmath,amssymb}
\usepackage{qic}

\usepackage{graphicx}
\usepackage{dcolumn}
\usepackage{bm, color}
\usepackage{subfig}
\usepackage{algpseudocode}
\usepackage{algorithmicx}
\usepackage{algorithm}

\renewcommand{\epsilon}{\varepsilon}

\newcommand{\polylog}{\textrm{polylog}}

\renewcommand{\dh}{\ensuremath{\mathcal{D(H)}}}

\renewcommand{\b}{\ensuremath{\mathcal{B}}}

\newcommand{\e}{\ensuremath{\mathcal{E}}}
\newcommand{\f}{\ensuremath{\mathcal{F}}}

\newcommand{\h}{\ensuremath{\mathcal{H}}}

 \renewcommand{\t}{\ensuremath{\mathcal{T}}}

\renewcommand{\o}{\ensuremath{\mathcal{O}}}

 \newcommand{\tr}{{\rm tr}} 
 
 \newcommand{\supp}[1]{\mathit{supp}( #1 )}

\newcommand{\bra}[1]{\langle #1 |}
\newcommand{\ket}[1]{| #1 \rangle}
\newcommand{\braket}[2]{\langle #1 | #2 \rangle}
\newcommand{\ketbra}[2]{| #1 \rangle \langle #2 |}

\begin{document}

\setlength{\textheight}{8.0truein}    
\runninghead{An HHL-Based  Algorithm for Computing Hitting Probabilities of Quantum Walks}
            {Ji Guan, Qisheng Wang, Mingsheng Ying}
\normalsize\textlineskip
\thispagestyle{empty}
\setcounter{page}{1}

\copyrightheading{0}{0}{2003}{000--000}

\vspace*{0.88truein}

\alphfootnote

\fpage{1}

\centerline{\bf
An HHL-Based  Algorithm for Computing Hitting Probabilities of Quantum Walks}

\vspace*{0.37truein}

\centerline{\footnotesize 
Ji Guan\footnote{Corresponding author. Email: guanji1992@gmail.com.}}
\vspace*{0.015truein}

\centerline{\footnotesize\it State Key Laboratory of Computer Science, Institute of Software, Chinese Academy of Sciences, Beijing 100190, China}
\baselineskip=10pt
\vspace*{10pt}
\centerline{\footnotesize
Qisheng Wang
}
\vspace*{0.015truein}
\centerline{\footnotesize\it Department of Computer Science and Technology, Tsinghua University, Beijing 100084, China}
\baselineskip=10pt
\vspace*{10pt}
\centerline{\footnotesize 
Mingsheng Ying}
\vspace*{0.015truein}
\centerline{\footnotesize\it Centre for Quantum Software and Information,
University of Technology Sydney, NSW 2007, Australia}
\baselineskip=10pt
\centerline{\footnotesize\it State Key Laboratory of Computer Science, Institute of Software, Chinese Academy of Sciences, Beijing 100190, China}
\baselineskip=10pt
\centerline{\footnotesize\it Department of Computer Science and Technology, Tsinghua University, Beijing 100084, China}
\vspace*{0.225truein}
\publisher{(received date)}{(revised date)}

\vspace*{0.21truein}


\abstracts{
We present a novel application of the HHL (Harrow-Hassidim-Lloyd) algorithm --- a quantum algorithm solving systems of linear equations --- in solving an open problem about quantum walks, namely computing hitting (or absorption) probabilities of a general (not only Hadamard)  one-dimensional quantum walks with two absorbing boundaries. This is achieved by a simple observation that the problem of computing hitting probabilities of quantum walks can be reduced to inverting a matrix. Then a quantum algorithm with the HHL algorithm as a subroutine is developed for solving the problem, which is faster than the known classical algorithms by numerical experiments. }{}{}

\vspace*{10pt}

\keywords{Quantum walks, The HHL algorithm, Hitting probabilities}
\vspace*{3pt}
\communicate{to be filled by the Editorial}

\vspace*{1pt}\textlineskip    
\section{Introduction}
Quantum walks are a quantum counterpart of classical random walks~\cite{kempe2003quantum,venegas2012quantum}. Targeting applications in quantum optics, the first model of quantum walks was proposed by Aharonov et. al.~\cite{aharonov1993quantum} in 1993. After that, many different models of and proposals for implementing quantum walks were made (e.g.~\cite{kempe2003quantum,travaglione2002implementing,sanders2003quantum,dur2002quantum,zahringer2010realization,goyal2013implementing,schreiber20122d}). In the field of quantum computing, quantum walks have been employed to develop quantum algorithms beyond classical algorithms~\cite{ambainis2003quantum}, even achieving exponential speedup~\cite{childs2003exponential}.

Due to the fact that quantum walks exhibit interference patterns whereas classical random walks do not, quantum walks behave very differently from their classical counterparts~\cite{kempe2003quantum,venegas2012quantum}.  Hitting (or absorption) probability is one of the earliest evidence of this difference~\cite{kempe2003quantum}.
On the other hand, the hitting probabilities of quantum walks are also an important issue in analyzing the quantum walk-based algorithms; in particular, their termination probability and expected running time.

The hitting probabilities of one-dimensional quantum walks have been studied in a series of papers. 
If an unbiased (Hadamard) one-dimensional quantum walk with two absorbing boundaries starts from the position next to one boundary, then the hitting probability distribution over the two boundaries is approaching $(1/\sqrt{2}, 1-1/\sqrt{2})$ when the number of positions tends to $\infty$, while the corresponding value is $(1,0)$ in the classical case~\cite{ambainis2001one}. Given a fixed number of positions, the hitting probabilities of an unbiased quantum walk were considered in~\cite{Bach2004} for the case where the walk starts very far from one barrier but an arbitrary
distance from the other barrier. Some efforts have been also made to compute the hitting probabilities of a one-dimensional quantum walk with a computational basis state as the initial state (e.g.~\cite{yamasaki2003analysis,konno2002absorption,bach2009absorption,kwek2011one,kuklinski2018absorption,kuklinski2020conditional}). Notably, an explicit form of the hitting probabilities for any number of positions has been obtained in the case of unbiased one-dimensional quantum walk~\cite{bach2009absorption}. 
However, the problem of computing the hitting probabilities of a one-dimensional quantum walk in the most general case --- possibly biased and starting from a superposition or entanglement of positions and directions --- is still unsolved.

In this work, we present a quantum algorithm for computing the hitting probabilities of a one-dimensional general quantum walk. Our algorithm is led by the observation that given an initial state, computing the hitting probabilities of a one-dimensional quantum walk with two absorbing boundaries can be reduced to the problem of inverting a matrix. Thus, the HHL algorithm solving systems of linear equations can be employed as the main subroutine. Note that under certain assumptions, the HHL algorithm can be exponentially faster than classical algorithms for the same purpose~\cite{harrow2009quantum}. Taking this advantage of the HHL algorithm and numerical experiments, the quantum algorithm for computing hitting probabilities of quantum walks given in this paper is faster than the known classical algorithms.

The paper is organized as follows. Section~\ref{sec:walk} reviews the setting of one-dimensional general quantum walks. Section~\ref{sec:hitting} shows how can computing hitting probabilities of quantum walks be reduced to inverting a matrix. In Section~\ref{sec:algorithm}, we develop a quantum algorithm for computing the hitting probabilities. A classical algorithm for the same purpose is also given there so that a clear comparison between the quantum and classical algorithms can be made. Brief discussion and conclusion are drawn in Section~\ref{sec:discuss} and~\ref{sec:conclusion}, respectively.

\section{One-Dimensional Quantum Walks}\label{sec:walk}
 First, we recall the basic setting~\cite{Bach2004} of a general quantum  walk on a one-dimensional lattice indexed by integers $0$ to $n$. Assume that two absorbing boundaries are at positions $0$ and $n$, respectively. Let $\h_d$ be the direction space, which is a 2-dimensional Hilbert space with orthogonal basis states $\ket{L}$ and $\ket{R}$, indicating directions left and right, respectively. Let $\h_p$ be an $(n+1)$-dimensional Hilbert space with orthonormal basis states $\ket{0}, \ket{1},\ldots, \ket{n}$, where the vector $\ket{k}$ is used  to denote position $k$ for each $0\leq k\leq n$. Thus, the state space of the quantum  walk is $\h=\h_p \otimes \h_d$. Each step of the walk consists of the following three sub-steps:
\begin{itemize}
    \item Measure the position of the system to see whether the current position is 0 or $n$ (boundary). If the outcome is ``yes'', then the walk terminates; otherwise, it continues. Mathematically, the measurement can be described by
$$\{M_{yes} =(\ket{0}\bra{0}+\ket{n}\bra{n})\otimes I_d, M_{no} =I-M_{yes}\},$$
where $I_d$ and $I$ are the identity operators on $\h_d$ and $\h$, respectively.
\item  A ``coin-tossing'' operator
\begin{equation}\label{Eq_tossing}
    T=\left(\begin{matrix}
    a&b\\
    -e^{i\theta}b^*&e^{i\theta}a^*
\end{matrix}\right)=\ket{L}\bra{\top}+\ket{R}\bra{\bot}
\end{equation}
is applied on the direction space, where $a,b$ are complex numbers satisfying the normalization condition: $|a|^2 + |b|^2 = 1$, $\theta$ is a real number, \begin{equation}\label{2ket}\ket{\top}= (a^*\ket{L}+ b^*\ket{R}),\  \ket{\bot}= e^{-i\theta}(-b\ket{L}+a\ket{R})\end{equation} and $a^*$ is the conjugation of complex number $a$.  Note that the coin tossing operator here is a general $2\times2$ unitary operator rather than the special Hadamard gate with $a=b=1/\sqrt{2}$ and $\theta=\pi$ considered in~\cite{ambainis2001one}. In the following, we assume that $a\not=0$ and  $b\not=0$; otherwise, the walk will only move to one direction and the randomness is degenerated to the determinism.
 \item  A shift operator
 \[S=\sum_{k=0}^{n}\ket{k\ominus1}\bra{k}\otimes\ket{L}\bra{L}+\ket{k\oplus 1}\bra{k}\otimes \ket{R}\bra{R}\]
is performed on the space $\h$. The intuitive meaning of the operator $S$ is that the system walks one step left or right according to the direction state. Here, $\oplus$ and $\ominus$ stand for addition and subtraction modulo $n+1$, respectively.
\end{itemize} 
Combining the above last two sub-steps, the purely quantum evolution of the walk without the measurement is described as  
\begin{align*}
    U&=S (I_p\otimes T)\\
    &=\sum_{k=0}^{n}\ket{k\ominus1}\bra{k}\otimes\ket{L}\bra{\top}+\ket{k\oplus 1}\bra{k}\otimes \ket{R}\bra{\bot},
\end{align*}
where $I_p$ is the identity operator on $\h_p$.

A quantum walk starts from an initial state $\ket{\psi_0}$. In previous works, the initial state $\ket{\psi_0}$ is usually predefined and chosen as a special  basis state of $\h$ (e.g. $\ket{\psi_0}=\ket{1}_p\ket{L}_d$ in~\cite{ambainis2001one}). In this paper, we consider the most general case with the initial state $\ket{\psi_0}$ being quickly prepared by an oracle (a block box of a quantum circuit) $\o$ within time $O(\polylog(n))$ (note that if we use $\lceil\log (n+1)\rceil$ qubits to encode the $(n+1)$ positions, then the time is polynomial in the number of qubits). Formally,
\begin{equation}\label{eq:oracle}
    \ket{\psi_0}=\o\ket{1}_p\ket{L}_d
\end{equation}
 can be an arbitrary state in  $\h_p'\otimes\h_d$, where $\h_p'$ is the linear space spanned by $\{\ket{1},\ldots,\ket{n-1}\}$, excluding the absorbing positions $\ket{0}$ and $\ket{n}$. Here, we assume that the walk does not start from the absorbing positions $0$ and $n$ because otherwise, its behavior is trivial; that is, the walk will stay there forever. 

\section{Hitting Probabilities}\label{sec:hitting}

The hitting problem is important in understanding the behaviors of classical random walks and in the analysis of random walk-based algorithms. It plays a similar role in quantum walks~\cite{venegas2012quantum,kuklinski2018absorption}. Given a  quantum walk and an initial state,  the hitting probability is defined as the probability that the walk hits the boundary (position $0$ or $n$) accumulated in a run of an infinite number of steps.

In this section, we show that computing the hitting probabilities of a quantum walk can be reduced to inverting a matrix.
First, we observe that a single step of the quantum walk, as described in the last section, can be modelled by the following super-operator (i.e. a completely positive and trace-preserving  map ~\cite{nielsen2010quantum}) $\e$ on $\h$:
\begin{equation}\label{eq:superoperator}
    \e(\rho)=UM_{no}\rho M_{no}^\dagger U^\dagger +M_{yes}\rho M_{yes}^\dagger, \ \forall \rho \in \dh,
\end{equation}
where $\rho$ is a mixed state (a semi-definite positive matrix on $\h$ with trace unit $\tr(\rho)=1$), and $\dh$ denotes the set of all mixed states on $\h$. 
Noting that $M_{no}M_{yes}=0$, one can check by induction on $m$ that 
the probability of the walker being at position $k$ after $m$ steps is 
\[\tr(P_{k}\e^{m}(\ketbra{\psi_0}{\psi_0})),\]
where $\{P_k=\ketbra{k}{k}\otimes I_{d}\}_{k=0}^{n}$ is a projective measurement. Subsequently, the hitting probabilities at positions $0$ and $n$ are  
\begin{align}
    &\lim_{m\rightarrow \infty}\tr(P_{0}\e^{m}(\ketbra{\psi_0}{\psi_0})\label{Eq:position_0},\\ &\lim_{m\rightarrow \infty}\tr(P_{n}\e^{m}(\ketbra{\psi_0}{\psi_0})\label{Eq:position_n}, 
\end{align} respectively.

It has been shown in the previous literature~\cite{guan2018decomposition,Wang2016,Baumgartner2011,ying2013reachability,Albert2018} that with respect to a super-operator $\e$, the state space $\h$ can be decomposed into the finite direct sum of mutually orthogonal minimal subspaces (see Definition \ref{minimal-s} in Appendix \ref{proof}) together with the maximum transient subspace $\t\subseteq \h$, which is orthogonal to all of the minimal subspaces.  
Moreover, an efficient algorithm for this decomposition was developed (see~e.g.  \cite{guan2018decomposition,ying2013reachability}). One essential fact  here is the transiency of $\t$ that starting from any initial state $\rho\in\dh$, the state will be eventually absorbed into the  minimal subspaces~\cite{Baumgartner2011,ying2013reachability}:
\begin{equation}\label{Eq:transient}
    \lim_{m\rightarrow \infty}\tr((I-P_\t)\e^{m}(\rho))=1,
\end{equation}
where $P_\t$ is the projection onto $\t$ and $I$ is the identity operator on $\h$. 

Applying this decomposition technique to the super-operator $\e$ of the quantum  walk defined in Eq.~(\ref{eq:superoperator})  yields:
\begin{equation}\label{Eq:decomposition}
    \h=\h_p\otimes\h_d=(\oplus_{k=1}^4 \b_k)\oplus\t,
\end{equation}
where $\t$ is the transient subspace and the four minimal subspaces $\{\b_{k}\}^{4}_{k=1}$ are one-dimensional subspaces linearly spanned respectively by the following pure states:
\[\ket{0}_p\ket{L}_d,\ket{0}_p\ket{R}_d,\ket{n}_p\ket{L}_d,\ket{n}_p\ket{R}_{d}.\]

Then the probabilities of hitting the four minimal subspaces are:\begin{equation}\label{Eq:eventual_prob}
    p_k=\lim_{m\rightarrow \infty}\tr(P_{\b_k}\e^{m}(\ketbra{\psi_0}{\psi_0})), \ k=1,2,3,4.
\end{equation}
 From Eqs. (\ref{Eq:position_0}), (\ref{Eq:position_n}) and (\ref{Eq:eventual_prob}), we see that $p_1+p_2$ and $p_3+p_4$ are the hitting probabilities of the quantum  walk at positions $0$ and $n$, respectively. Moreover, it is easy to see that the walker can reach the left (resp. right) boundary only from  the left (resp. right) direction. Thus, $p_2=p_3=0$.  At the same time, by Eq.(\ref{Eq:transient}) and the trace-preserving property of $\e$, we have $p_1+p_2+p_3+p_4=1$.  Therefore, $p_2=1-p_1$, and we only need compute $p_1$. For simplicity, we write $p$ for $p_1$ in the following discussion.
 
 Now we can present the main result of this section showing that $p$ can be computed through inverting a matrix. 
 
 \begin{theorem}\label{Theo:hitting}
     Given a quantum walk as defined in Section~\ref{sec:walk} and an initial state $\ket{\psi_0}\in \h_p'\otimes \h_d$, the hitting probability at position $0$  is 
\begin{equation}\label{Eq:p}
    p=\bra{1,\top,1,\top^*}(I-M_{\e_{t}})^{-1}\ket{\psi_0,\psi_0^*},
\end{equation}
 where $\ket{\psi^*}$ is the entry-wise conjugation of $\ket{\psi}$, $M_{\e_t}=M\otimes M^*$ with 
\[M=\sum_{k=2}^{n-1}\ket{k-1}\bra{k}\otimes\ket{L}\bra{\top}+\sum_{k=1}^{n-2}\ket{k+ 1}\bra{k}\otimes \ket{R}\bra{\bot}, 
\] 
and $\ket{\top}, \ket{\bot}$ being given in Eq.(\ref{2ket}).
\end{theorem}

For readability, we postpone the proof of the above theorem into Appendix \ref{proof}.

\section{A Quantum Algorithm for Computing the Hitting Probabilities}\label{sec:algorithm}
In this section, we develop a quantum algorithm for computing the hitting probability $p$ based on Theorem~\ref{Theo:hitting}. At the end of this section, we will compare it with a classical algorithm for the same purpose.

Our quantum algorithm will use the HHL algorithm as the main subroutine. The HHL algorithm was developed by Harrow, Hassidim, and Lloyd~\cite{harrow2009quantum} for solving the quantum linear system problem (QLSP): \begin{itemize}\item Given an $N$-by-$N$ matrix  $A$ (whose elements are accessed by an oracle)
 and a quantum state $\ket{b}$, find a quantum state  $\ket{x}$ and a normalization factor $\mu$  such that $\mu A\ket{x}=\ket{b}$.\end{itemize} 
 It should be emphasized that successfully preparing $\ket{b}$ is subject to the resource for state preparations. In general, one cannot expect to obtain an arbitrary state $\ket{b}$, and thus initializing a QLSP is an essential issue before solving it by the HHL algorithm in practical applications~\cite{clader2013preconditioned}. As we will see later, this issue arises in our goal of computing hitting probabilities using the HHL algorithm, and need to be resolved by a certain limitation on initial states.
 
 Obviously, QLSP is a quantum analog of the linear system problem (LSP) in the classical world,  a common practical problem that arises both on its own and as a subroutine in more complex problems: \begin{itemize}\item Given a $N$-by-$N$ matrix $A$
 and a vector $\vec{b}$, find a vector $\vec{x}$ such that $A\vec{x}=\vec{b}$.\end{itemize} The HHL algorithm can   exponentially speed up the best classical method for solving LSP  under the following four constraints~\cite{harrow2009quantum,aaronson2015read}: 
\begin{enumerate}
    \item  $\ket{b}$  can be prepared quickly to load the information of $\vec{b}$;
    \item the matrix $A$ must be  $s$-sparse (that is, $A$ has at most $s$ nonzero entries per row) for some constant number $s$, or it can be efficiently decomposed into $s$-sparse sub-matrices;
    \item $A$ is well-conditioned in the sense that the condition number $\kappa$ of $A$ (the ratio between $A$'s maximal and minimal singular values) must scale as $O(\polylog (N))$;
    \item a limited statistical information about $\vec{x}$   is the targeting goal instead of the output $\vec{x}$ itself: for example, the approximate value of an inner product $\vec{y}^\dagger\vec{x}$ for a given vector $\vec{y}$.
\end{enumerate}
    In this context, the total complexity of the HHL algorithm in solving LSP is $\tilde O(\kappa^2 \log N)$, where $\tilde O(\cdot)$ suppresses poly-logarithmic factors of $\kappa$ and $\log N$, while the best classical counterpart is $O(N\sqrt{\kappa})$~\cite{harrow2009quantum}. The exponential speedup is achieved as long as $\kappa=O(\polylog (N))$.
However, in practice, systems of linear equations with a polylogarithmic condition number are quite rare~\cite{brenner2007mathematical,bank1989conditioning}. It is much more common for a system to have a condition number that scales as polynomials in $N$. Fortunately,  the condition number dependence of the HHL algorithm was significantly improved by Ambainis \cite{ambainis2012variable} from $\kappa^2$ to $\kappa\log^3\kappa$. Consequently, at least a polynomial speedup can be demonstrated if $\kappa=O(N^c)$ for $c<2$. 

The quantum algorithm HHL has been employed in solving a series of problems from the classical world, including the calculation of electromagnetic scattering cross-sections for the systems involving smooth geometric figures in 3-dimensional space~\cite{clader2013preconditioned}, solving large systems of differential equations~\cite{leyton2008quantum,berry2014high}, data fitting~\cite{wiebe2012quantum}, various tasks in machine learning~\cite{lloyd2013quantum} and approximating effective resistances in electrical networks~\cite{wang2013quantum}. 

Now we show how the HHL algorithm can be naturally applied in computing the hitting probabilities of quantum walks, a problem in the quantum world. By Theorem~\ref{Theo:hitting}, we see that this problem can be reduced to computing the probability: 
\[p=\mu\braket{1,\top,1,\top^*}{x},\]
where $\ket{x}$ with a normalization factor $\mu$ is the solution of the following QLSP: \begin{equation}\label{eq:QLSP}
\mu(I-M_{\e_t})\ket{x}=\ket{\psi_0,\psi_0^*},
\end{equation} and the dimension $N$ of $I-M_{\e_t}$ is $N=(2n-2)^2$.
Then our quantum algorithm for computing $p$ is presented as Algorithm~\ref{Alg:quantum}, which uses the HHL algorithm as a key subroutine.
\begin{algorithm}[htp]
\caption{Q-HittingProb($\o$)}
\label{Alg:quantum}
    \begin{algorithmic}[1]
    \Require An oracle $\o$ defined in Eq.(\ref{eq:oracle}) producing an initial state
    \Ensure The hitting probability  $p$ 
    \State\label{line:oracle}Call oracle $\o$ twice to produce two copies of the initial state $\ket{\psi_0,\psi_0}$, where $\ket{\psi_0}$ is the initial state;
    \State\label{line:HHL}Obtain $\ket{x}$ and its normalization factor $\mu$ by solving $\mu(I-M_{\e_t})\ket{x}=\ket{\psi_0,\psi_0}$ with the HHL algorithm;
    \State Prepare one qubit state $\ket{\top}$, $\ket{\top^*}$ and computational basis state $\ket{1}$;
    \State\label{line:swap}\Return $p=\mu\braket{1,\top,1,\top^*}{x}$ by performing SWAP test between  $\ket{1,\top,1,\top^*}$ and $\ket{x}$~\cite{buhrman2001quantum}.
    \end{algorithmic}  
\end{algorithm}

The design idea of Algorithm~\ref{Alg:quantum} deserves some careful explanations: 
\begin{enumerate}\item  The state preparation for initializing the QLSP requires that $\ket{\psi_0^*}$ can be obtained by calling the oracle $\o$ defined in Eq.(\ref{eq:oracle}). Thus we have to add a restriction on the initial state so that $\ket{\psi_0^*}=\ket{\psi_0}$; that is, all of the amplitudes of  $\ket{\psi_0}$ are real numbers.
Then in line~\ref{line:oracle}, we call the oracle $\o$ defined in Eq.(\ref{eq:oracle}) twice to get two copies of the initial sate $\ket{\psi_0}^{\otimes 2}=\ket{\psi_0,\psi_0}$ with time complexity $O(\polylog(n))$. 
\item In line~\ref{line:HHL}, state $\ket{\psi_0,\psi_0}$ and matrix $I-M_{\e_t}$ are fed into the HHL algorithm, and QLSP $\mu(I-M_{\e_t})\ket{x}=\ket{\psi_0,\psi_0}$ is solved with the solution state $\ket{x}$ and its normalization $\mu$. 
\item In line 3, we prepare one qubit states  $\ket{\top}$ and $\ket{\top^*}$ and a computational basis state $\ket{1}$, which can all be obtained in time $O(1)$.
\item In the last step (line~\ref{line:swap}), we apply the SWAP test~\cite{buhrman2001quantum} between $\ket{1,\top,1,\top^*}$ and $\ket{x}$ to compute inner product $\braket{1,\top,1,\top^*}{x}$, and then probability  $p=\mu\braket{1,\top,1,\top^*}{x}$ is obtained. This only needs a constant number of copies of these two states. \end{enumerate}

Let us further analyze the complexity of Algorithm~\ref{Alg:quantum}. First, we note that $I-M_{\e_t}$ is 5-sparse. The above analysis indicates that all of the four constraints for the HHL algorithm mentioned before are satisfied. Thus the total complexity of Algorithm~\ref{Alg:quantum} is $O(\kappa^2\polylog(n))$, where $\kappa$ is the condition number of matrix $I-M_{\e_t}$ defined in Theorem~\ref{Theo:hitting}, and $n+1$ is the number of the positions of the quantum walk defined  in Section~\ref{sec:walk}. For determining $\kappa$, we implement numerical experiments (see Appendix \ref{app-numeric}) and the result shows that $\kappa=O(n^{2.5})$ for any fixed parameters $a,b$ and $\theta$ of $M_{\e_{t}}$
with
$|a|\geq 1/\sqrt{2}$. In this case, the complexity of Algorithm~\ref{Alg:quantum} is  $O(n^5\polylog(n))$. If we replace the standard HHL algorithm by Ambainis' improved version~\cite{ambainis2012variable} in Algorithm~\ref{Alg:quantum}, then  the complexity can be reduced to $O(n^{2.5}\polylog (n)).$

For a better understanding about the advantage of our quantum algorithm, we present a classical algorithm for computing hitting probability $p$ as Algorithm~\ref{Alg:classical}.
 \begin{algorithm}[htp]
\caption{C-HittingProb($\o$)}
\label{Alg:classical}
    \begin{algorithmic}[1]
    \Require  An oracle $\o$ defined in Eq.(\ref{eq:oracle}) producing an initial state 
    \Ensure The hitting probability  $p$ 
    \State\label{line:c_oracle}Call oracle $\o$  to produce the initial state $\ket{\psi_0}$;
    \State\label{line:c_vector}Get vector $\vec{\psi_0}$ by measuring $\ket{\psi_0}$, where the amplitudes of $\ket{\psi_0}$ are the elements of $\vec{\psi_0}$;
    \State\label{line:C_solve}Solve LSP $(I-M_{\e_t})\vec{x}=\vec{\psi_0}\otimes \vec{\psi_0}$ to obtain $\vec{x}$
    \State\label{line:C_result}\Return $p=\vec{y}^\dagger\vec{x}$, where $\vec{y}$ is the vector form of $\ket{1,\top,1,\top^*}$.
    \end{algorithmic}  
\end{algorithm}

This algorithm has a design idea similar to that of Algorithm~\ref{Alg:quantum}. It also requires that we are able to call the oracle $\o$ in Eq.(\ref{eq:oracle}) and access the initial state $\ket{\psi_0}$. 
Its first step (line~\ref{line:c_oracle}) is to call oracle $\o$ producing the initial state $\ket{\psi_0}
$. The next step (line~\ref{line:c_vector}) is to write down vector $\vec{\psi_0}$ with the amplitudes of $\ket{\psi_0}$ as its elements, which requires $O(n)$ steps. In line~\ref{line:C_solve},  we solve LSP $(I-M_{\e_t})\vec{x}=\vec{\psi_0}\otimes \vec{\psi_0}$ rather than a QLSP. The last step (line~\ref{line:C_result}) is to compute the inner product $\vec{y}^\dagger\vec{x}$, where $\vec{y}$ is the vector form of known quantum state $\ket{1,\top,1,\top^*}$.
The best known way of finishing  the last two steps is to apply  conjugate gradient method  with time complexity $O(N\sqrt{\kappa}) = O(n^2\sqrt{\kappa})$~\cite{harrow2009quantum}, where $\kappa$ is the condition number of $I-M_{\e_t}$. Thus, the total complexity of Algorithm~\ref{Alg:classical} is $O(n^2\sqrt{\kappa})$. Again, for any fixed parameters $a$, $b$ and $\theta$ with $|a|\geq 1/\sqrt{2}$, the numerical result shows that $\kappa=O(n^{2.5})$, so the complexity of Algorithm~\ref{Alg:classical} is $O(n^{3.25})$. Therefore, according to the numeric analysis,  our quantum algorithm is faster than the classical one in the case of  $|a|\geq 1/\sqrt{2}$. 

We finally remark that as the main proof technique (see Lemma~\ref{Lem} in Appendix~\ref{proof}) of Theorem~\ref{Theo:hitting} does not depend on the topological structure of quantum walks, our results given in this section for one-dimensional quantum walks can be straightforwardly generalized to quantum walks on graphs~\cite{aharonov2001quantum}.

\section{Discussion}\label{sec:discuss}
In the above, we developed an HHL-based quantum algorithm to compute the hitting probabilities of one-dimensional quantum walks, which beats a classical algorithm achieving the same task in terms of the runtime for a modest size of walks.
An important question is whether the complexity of Algorithm~\ref{Alg:quantum} can be further improved? One possibility is to reduce the condition number $\kappa$ of $I-M_{\e_t}$. By calling a precondition oracle, we may achieve an exponential speedup as long as we find a certain sparse pattern of a matrix $B$ such that $B(I-M_{\e_t})$ has a constant condition number~\cite{clader2013preconditioned}. In this case, the complexity of Algorithm~\ref{Alg:quantum} is then $O(\polylog(N))$. 

On the other hand, an interesting question left open is: is there a faster classical algorithm that erodes the (exponential) speedup brought by the HHL algorithm (with a certain precondition)?

Recently, a quantum-inspired classical algorithm was proposed to solve LSP for low-rank matrices and achieved an exponential speed-up over the previous classical algorithms~\cite{gilyen2018quantum}. In particular, assuming length-square sampling access to input data, one can implement the pseudo-inverse of a low-rank matrix and statistically estimate the solution to the $N$-dimensional LSP problem $A\vec{x} = \vec{b}$ which is the same as the case in the HHL algorithm. The proven complexity of this classical algorithm is $\tilde O(\kappa^{16}r^6\|A\|_{F}^{6}\cdot\polylog(N))$, where $\kappa$ and $r$ are the condition number and rank of  $A$, respectively, and $\|A\|_{F}$ is the Frobenius norm of $A$ (the square root of the summation of the squares of singular values of $A$). As we can see, this quantum-inspired algorithm works for general low-rank matrix $A$, whereas the HHL algorithm exhibits an exponential speedup over all known classical algorithms for sparse, full-rank matrix $A$ because the time complexity of it is $\tilde O(\kappa^2\log N)$, which is independent on rank $r$.  Actually, other quantum-inspired classical algorithms in machine learning also only efficiently effect on low-rank matrices~\cite{arrazola2019quantum}, such as algorithms for recommendation systems~\cite{tang2019quantum}, principal component analysis~\cite{tang2018quantum} and support vector machine~\cite{ding2019quantum}. It seems impossible to efficiently apply these algorithms on high-rank matrices, as, if so, then classical computers can efficiently simulate quantum computers~\cite{harrow2009quantum}, i.e., $\textbf{BQP}=\textbf{P}$, which is strongly conjectured to be false.

In our application of the HHL algorithm in quantum walks, it is worth noting that $A=I-M_{\e_{t}}$ in Eq.(\ref{eq:QLSP}) of QLSP is a full-rank ($r=N$) matrix as $I-M_{\e_{t}}$ is invertible by Eq.(\ref{Eq:p}) in Theorem~\ref{Theo:hitting}. Thus a quantum-inspired algorithm as those discussed above cannot erase the speedup powered by the HHL algorithm in our case. Furthermore, other cleverly designed classical algorithms are also unlikely able to estimate the hitting probabilities of one-dimensional quantum walks without obtaining the classical information (all amplitudes) of the unknown initial quantum state $\ket{\psi_0}$, which requires at least time $O(n)$. This hints that possibly, our quantum algorithm together with a precondition oracle exponentially speeds up any classical algorithm for the same purpose.

\section{Conclusion}\label{sec:conclusion}
In this paper, we used the HHL algorithm as a subroutine to develop a quantum algorithm (Algorithm~\ref{Alg:quantum}) for computing the hitting probabilities of one-dimensional quantum walks. To the best of our knowledge, this is the first quantum algorithm designed for solving a problem about quantum walks. It was shown by the numerical experiment that a corresponding classical algorithm is much slower than the quantum algorithm.

\section*{Acknowledgments}
The authors are grateful to Professor Andris Ambainis for helpful discussions. This work was partly supported by the National Key R\&D Program of China 
(Grant No: 2018YFA0306701), the National Natural Science Foundation of China (Grant No: 61832015) and the Australian Research Council (ARC) under grant No. DP210102449.
\section*{References}
\bibliographystyle{unsrt}
\bibliography{note110418.bib}
\clearpage
\section*{Appendix}
\noindent
\section{Proof of Theorem 1}\label{proof}

Before presenting the proof of Theorem 1, we need some technical preparations. First, let us recall the notion of the minimal subspace from~\cite{Baumgartner2011,ying2013reachability}. 

\begin{definition}\label{minimal-s}
    Given a super-operator $\f$ on $\h$, 
   \begin{itemize}
        \item a state $\rho$ is called a stationary state if $\f(\rho)=\rho$; furthermore, $\rho$ is minimal if there is no other stationary state $\sigma$ with $\supp{\sigma}\subseteq\supp{\rho}$;
      \item a subspace $\h'$ of  $\h$ is called a minimal  subspace if it is a support of a minimal stationary state,   
    \end{itemize}
    where the support of $\rho$ is the subspace of $\h$ linearly spanned by the eigenvectors corresponding to non-zero eigenvalues of $\rho$. \end{definition}

Given a completely positive map $\f$ on $\h$. Then $\f$ admits a representation as $$\f(A)=\sum_{k}F_{k}A F_{k}^\dagger,$$ where operators $\{F_k\}_k$ on $\h$ are called the Kraus operators of $\f$~\cite{nielsen2010quantum}. In the following discussions, we use $\{F_k\}_k$ to denote $\f$ as $\f=\{F_k\}_k$.   Furthermore, the matrix representation of $\f$  is defined as ~\cite{ying2016foundations,wolf2012quantum}:
\[M_\f=\sum_{k}F_k\otimes F_k^*,\]
where $F_k^*$ is the entry-wise conjugation of $F_k$.


Let $\b\subseteq \h$ be a minimal subspace under $\f$ and $\t$ the transient subspace. Then we define:
\begin{itemize}
    \item Shift operator: $$\f_{s}(\rho)=\sum_{k}P_\b F_{k}P_{\t}\rho P_\t F_k^\dagger P_{\b};$$
    \item Transient operator: $$\f_{t}(\rho)=\sum_{k}P_\t F_{k}P_{\t}\rho P_\t F_k^\dagger P_{\t}.$$
\end{itemize}
Intuitively, shift operator $\f_s$ represents the transferring effect of $\f$ from subspace $
\t$ to $\b$, and transient operator $\f_t$ is the restriction of $\f$ on $\t$.
The following lemma gives a way to compute the hitting probability of $\b$ by using these two operators.  

\begin{lemma}\label{Lem}
Let $\f$ be a super-operator on $\h$,  and $\b\subseteq \h$  a minimal subspace. Then for a given initial state $\rho\in\dh$, the hitting probability of $\b$ is
\[\lim_{m\rightarrow \infty}\tr(P_\b\f^{m}(\rho))=\tr(P_\b\rho)+\sum_{m=0}^{\infty}\tr(\f_s\circ\f^{m}_t(\rho)).\]
Furthermore, 
\[\sum_{m=0}^{\infty}\tr(\f_{s}\circ\f^{m}_t(\rho))=\bra{\Omega}M_{\f_s}(I-M_{
\f_{t}})^{-1}(\rho\otimes I)\ket{\Omega},\]
where $I$ is the identity operator on $\h$, and $\ket{\Omega}$ is the unnormalized maximum entangled state on $\h\otimes \h$, i.e., $\ket{\Omega}=\sum_{i}\ket{i}\otimes\ket{i}$ with an orthonormal basis $\{\ket{i}\}$ of $\h$.
\end{lemma}
{\it Proof.}
The Hilbert space $\h$ has a  minimal subspace decomposition:
\[\h=(\oplus_{j=1}^J\b_{j})\oplus\t.\]
Correspondingly, by the definition of minimal subspaces, we have the block matrix forms of Kraus operators $F_k$ of $\f$ (see more details in \cite{guan2018structure} ):

\[F_k=\left[\begin{matrix}
    F_{k,1}&\ldots&\ldots &T_{1}\\
    &\ddots & &\vdots\\
    &&F_{k,J}& T_{J}\\
    &&& T
\end{matrix}\right].\]
Similarly, any state $\rho$ has the following block matrix form:
\[\rho=\left[\begin{matrix}
    \rho_{1,1}&\rho_{1,2}&\ldots &\rho_{1,J+1}\\
    \vdots&\vdots &\ddots &\vdots\\
    \rho_{J+1,1}&\rho_{J+1,2}&\ldots& \rho_{J+1,J+1}
\end{matrix}\right].\]
Then by the multiplication rules  of block matrices, $\sum_{k}F_k^\dagger F_k = I$ (the trace-preserving property of $\f$) and induction on $m\geq 1$, we have:
\[\tr(P_{B_1}\f^m(\rho))=\tr(P_{\b_1}\rho)+\sum_{k=0}^{m}\tr(\f_s\circ\f^{k}_t(\rho)),\]
where $\f_s=\{P_{\b_1}F_kP_\t\}$ and $\f_t=\{P_{\t}F_kP_\t\}$.

For any matrix $A$ on $\h$, we have~\cite{wolf2012quantum}: $$\tr(A)=\bra{\Omega}A\otimes I\ket{\Omega},\quad A\otimes I\ket{\Omega}=I\otimes A^{T}\ket{\Omega}$$ where  $A^{T}$ is the transpose of $A$. Therefore, we obtain: 
  \[\tr(\f_{s}\circ\f^{m}_t(\rho))=\bra{\Omega}M_{\f_s}M_{
\f_{t}}^m(\rho\otimes I)\ket{\Omega}.\]
The absolute values of all eigenvalues of $M_{\f_{t}}$ are less than $1$, so $\lim_{m\rightarrow\infty} M_{\f_{t}}^m=0$. Thus, \[\sum_{k=0}^{\infty}M_{\f_{t}}^{m}=(I-M_{\f_{t}})^{-1},\] where $I$ is the identity operator on $\h\otimes \h$.  \hfill $\Box$

Now we are ready to prove  Theorem~\ref{Theo:hitting}. 

{\it Proof.}
Let us apply  Lemma~\ref{Lem} to the super-operator $\e$ of the quantum walk defined in Eq.(\ref{eq:superoperator}), minimal subspace $\b_1$ in Eq.(\ref{Eq:decomposition}) and the initial state $\ketbra{\psi_0}{\psi_0}$ in Eq.(\ref{eq:oracle}). Then we have:  \[p=\bra{\Omega }(\ketbra{1,\top}{1,\top}\otimes I)(I-M_{\e_{t}})^{-1}(\ketbra{\psi_0}{\psi_0}\otimes I)\ket{\Omega},\]
 where $\ket{\Omega}$ is the unnormalized maximum entangled state on $\h\otimes\h$ and $\h=\h_d'\otimes\h_p$ ($\h_p'$ is defined in the below of Eq.(\ref{eq:oracle})). 
Thus, the conclusion of Theorem~\ref{Theo:hitting} is obtained by noting that
\[(\ketbra{\psi}{\psi}\otimes I)\ket{\Omega}=\ket{\psi,\psi^*},\quad \forall \ket{\psi}\in \h.\]\hfill $\Box$

\section{Numerical Experiments of the Condition Number of $I-M_{\e_t}$}\label{app-numeric}
We developed a MATLAB program to randomly set the values of the parameters $\theta,a,b$ with $|a|+|b|=1$ and $|a|\geq 1/\sqrt{2}$ of ``coin-tossing'' operator $U$ defined in Eq.(\ref{Eq_tossing}). These parameters also appear in the matrix $I-M_{\e_t}$. We compute $\kappa$ for $n$ from 3 to 60, and the computation was done on a laptop. Totally, 1000 experiments have been done. One experiment result has shown in the following, and the others are similar. These computational results show that \begin{equation}\label{un-proved}\kappa(I-M_{\e_t})=O(n^{2.5}),\end{equation} where $n$ is the number of positions of quantum walks.   Unfortunately, we are unable to give a mathematical proof of the claim in Eq.(\ref{un-proved}).

\begin{figure}[H]
\centering
\includegraphics[width=3.5in,height=3in]{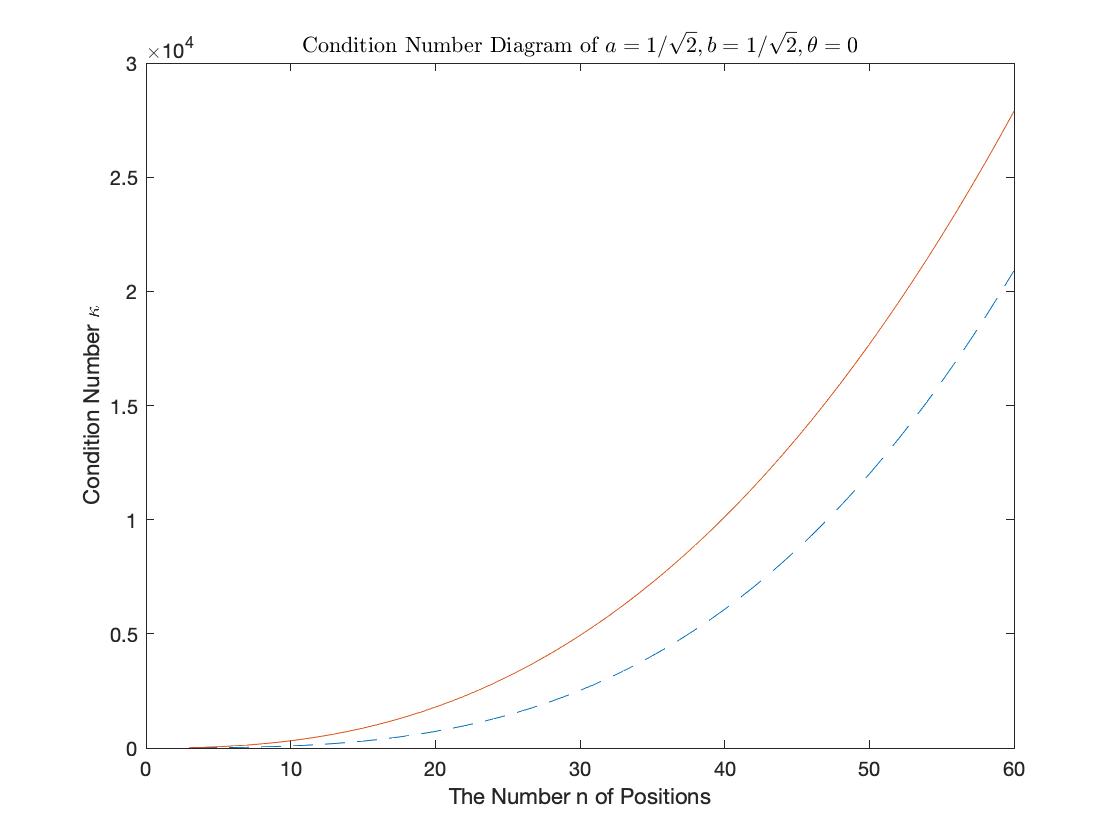}
\centering
\caption{Condition number diagram of $a=\frac{1}{\sqrt{2}},b=\frac{1}{\sqrt{2}}$ and $\theta=0$. The blue line represents the condition number of $I-M_{\e_t}$ and the red line is $n^{2.5}$.}
\label{fig:1}
\end{figure}
\end{document}